%% file: Mott_top_graphene.tex
\documentclass[aps,prl,twocolumn,superscriptaddress]{revtex4-2}
\usepackage{stix}

\usepackage[dvipsnames]{xcolor}
\definecolor{webblue}{RGB}{0,0,128}
\definecolor{darkpink}{HTML}{800040}
\definecolor{strongorange}{HTML}{BF5E26}
\usepackage[
hidelinks,
colorlinks=true,
citecolor=webblue,
linkcolor=strongorange,
urlcolor=darkpink
]{hyperref}

\usepackage{graphicx}
\usepackage{amsmath}
\usepackage{bm}
\usepackage{dsfont}
\usepackage[T1]{fontenc}
\usepackage[utf8]{inputenc}

\definecolor{amaranth}{HTML}{e52b50}

\newcommand{\bra}[1]{\langle#1\vert}
\newcommand{\ket}[1]{\vert#1\rangle}

\newcommand{\braket}[2]{\langle #1 \vert #2 \rangle}
\DeclareMathOperator{\Tr}{\mathrm{Tr}}

\usepackage{orcidlink}
\usepackage{floatrow}
\usepackage{revtex-supplement}

\newcommand{\prlsection}[1]{\textit{#1}.~}

\begin{document}

\title{Real-Space Switching of Local Moments Driven by Quantum Geometry in Correlated Graphene Heterostructures}

\author{Niklas Witt\,\orcidlink{0000-0002-2607-4986}}
\thanks{These two authors contributed equally.}
\affiliation{Institut f\"ur Theoretische Physik und Astrophysik and W\"urzburg-Dresden Cluster of Excellence ct.qmat, Universit\"at W\"urzburg, 97074 W\"urzburg, Germany}
\affiliation{I. Institute of Theoretical Physics,
	Universit\"at Hamburg,
        Notkestra{\ss}e 9,
	22607 Hamburg,
	Germany}
\affiliation{The Hamburg Centre for Ultrafast Imaging,
	Luruper Chaussee 149,
	22761 Hamburg,
	Germany}

\author{Siheon Ryee\,\orcidlink{0000-0002-5551-4223}}
\thanks{These two authors contributed equally.}
\affiliation{I. Institute of Theoretical Physics,
	Universit\"at Hamburg,
        Notkestra{\ss}e 9,
	22607 Hamburg,
	Germany}
\affiliation{The Hamburg Centre for Ultrafast Imaging,
	Luruper Chaussee 149,
	22761 Hamburg,
	Germany}    
 
\author{Lennart Klebl\,\orcidlink{0000-0002-5453-9779}}
\affiliation{Institut f\"ur Theoretische Physik und Astrophysik and W\"urzburg-Dresden Cluster of Excellence ct.qmat, Universit\"at W\"urzburg, 97074 W\"urzburg, Germany}
\affiliation{I. Institute of Theoretical Physics,
	Universit\"at Hamburg,
        Notkestra{\ss}e 9,
	22607 Hamburg,
	Germany}

\author{Jennifer Cano\,\orcidlink{0000-0003-1528-4344}}
\affiliation{Department of Physics and Astronomy, Stony Brook University, Stony Brook, New York 11794, USA}
\affiliation{Center for Computational Quantum Physics, Flatiron Institute, New York, New York 10010, USA}

\author{Giorgio Sangiovanni\,\orcidlink{0000-0003-2218-2901}}
\affiliation{Institut f\"ur Theoretische Physik und Astrophysik and W\"urzburg-Dresden Cluster of Excellence ct.qmat, Universit\"at W\"urzburg, 97074 W\"urzburg, Germany}

\author{Tim O. Wehling\,\orcidlink{0000-0002-5579-2231}}
\affiliation{I. Institute of Theoretical Physics,
	Universit\"at Hamburg,
        Notkestra{\ss}e 9,
	22607 Hamburg,
	Germany}
\affiliation{The Hamburg Centre for Ultrafast Imaging,
	Luruper Chaussee 149,
	22761 Hamburg,
	Germany}

\begin{abstract}
Graphene-based multilayer systems serve as versatile platforms for exploring the interplay between electron correlation and topology, thanks to distinctive low-energy bands marked by significant quantum metric and Berry curvature from graphene's Dirac bands. Here, we investigate Mott physics and local spin moments in Dirac bands hybridized with a flat band of localized orbitals in functionalized graphene. Via hybridization control, a topological transition is realized between two symmetry-distinct site-selective Mott states featuring local  moments in different Wyckoff positions, with a geometrically enforced metallic state emerging in between. We find that this geometrically controlled real-space switching of local moments and associated metal-insulator physics may be realized through proximity coupling of epitaxial graphene on SiC(0001) with group IV intercalants, where the Mott state faces geometrical obstruction in the large-hybridization limit. Our work shows that chemically functionalized graphene provides a correlated electron platform, very similar to the topological heavy fermions in graphene moiré systems but at significantly enhanced characteristic energy scales.
\end{abstract}

\maketitle

\makeatletter
\addtocontents{toc}{\string\tocdepth@munge}
\makeatother

Graphene, initially celebrated as a weakly interacting Dirac material, has developed towards a material basis for correlated electron physics---particularly in twisted graphene multilayers, where the moiré potential localizes parts of the electronic states to form flat bands featuring strong electron correlations~\cite{Balents2020, Nuckolls2024, cao2018correlated, cao2018unconventional, lu2019superconductors,  yankowitz2019tuning, sharpe2019emergent, nuckolls2020strongly, serlin2020intrinsic, shen2020correlated, hao2021electric, chen2021electrically,  park2021tunable, saito2021hofstadter,  wu2021chern, xie2021fractional,  rubio-verdu2022moire, Park2022, Zhang2022}. A seemingly complementary platform are moiré-less rhombohedral multilayers like Bernal bilayer, ABC trilayer, and ABCA tetralayer where correlations emerge in absence of localized electronic orbitals but from itinerant electrons near Van Hove singularities~\cite{Pantaleon2023, seiler2022quantum, tsui2024direct, zhang2023enhanced, zhou2021half, zhou2021superconductivity, zhou2022isospin, lu2024fractional, Holleis2025, Winterer2024, Kerelsky2021, Wirth2022}. As a common thread in both platforms, the low-energy states feature sizable Berry curvature and quantum metric, which are inherited from the graphene Dirac cones~\cite{wehling2014,Katsnelson2020,Liang2017,Marsal2024}. These quantum geometric contributions are suggested to play a decisive role in shaping emergent correlated states in these systems~\cite{Yu2025,Po2018,Song2022,Rai2024,Adak2024}, including the stabilization of superconducting order~\cite{Peotta2015,Toermae2022,Liang2017,Hu2019,Xie2020,Julku2020,HerzogArbeitman2022,Huhtinen2022,Yu2024,Chen2024,Hu2025,Tian2023,Tanaka2025} and various kinds of (pseudo)magnetic states~\cite{Song2019,Liu2021,Shimazaki2015,Wu2020,Tschirhart2021,Grover2022,Abouelkomsan2023}.

The emergent ordered states in moiré and rhombohedral multilayers set in at temperature scales of a few Kelvin or below, i.e., $\mathcal{O}$(meV)~\cite{Pantaleon2023,Nuckolls2024}. What sets these scales is an open matter. Considering the most plausible scenario for the moiré case from a strong-coupling perspective, the width of the flat bands ($\sim 10$\,meV) and the hybridization gap ($\sim 50$\,meV) determine order parameter stiffness, therefore also transition temperatures~\cite{Hu2019,Xie2020,Tian2023,Tanaka2025}. These energy scales are essentially governed by the graphene interlayer tunellings and moiré potential modulations. It is questionable whether van der Waals engineering can significantly increase them.

In this Letter, we propose an alternative material platform that exhibits strong quantum geometric effects and rich correlation phenomena, while also possessing a high intrinsic energy scale on the order of electronvolts. We first demonstrate how hybridization of Dirac bands with localized orbitals and electron interactions lead to flat bands that feature ``obstructed'' Mottness: depending on the hybridization strength, the atomic limit of the emergent site-selective Mott states corresponds to local moments forming in distinct Wyckoff positions as identified by the Luttinger surface. Thus, a hybridization-induced topological transition between two symmetry-distinct site-selective Mott states takes place with a protected metallic state manifesting in between. We further show that the aforementioned topological transition can be partly realized in real materials through proximity coupling of epitaxial graphene on SiC(0001) with group IV intercalants. 

\begin{figure}[!htbp]
    \centering
    \includegraphics[width=1.0\columnwidth]{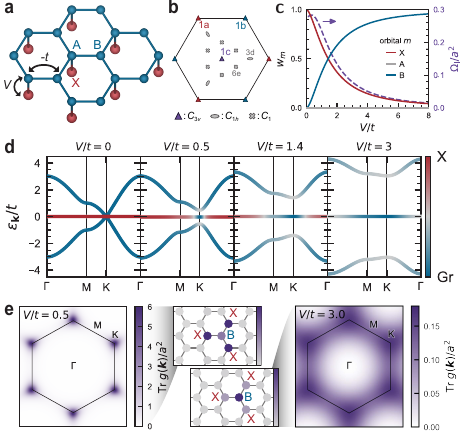}
    \caption{Flat band hybridized to graphene. (\textbf{a}) Lattice structure of decorated graphene honeycomb lattice with impurity X hybridized to sublattice site A created with VESTA~\cite{VESTA_Momma2011}. Only hopping $t$ between sublattices A and B as well as $V$ between X and A exist. (\textbf{b}) Wyckoff positions and their respective local point symmetry groups for the wallpaper group $p3m1$ (No.~156) representing the geometry in panel a. (\textbf{c}) Orbital weight $w_m = \sum_{\bm{k}} |w_{\bm{k}m}|^2/N_{\bm{k}}$ ($m\in\lbrace\mathrm{A,B,X}\rbrace$) with weight $|w_{\bm{k}m}|^2$ of the flat band crossing the Fermi level at $\bm{k}$, and minimal quadratic Wannier function spread $\Omega_{\mathrm{I}}$ as obtained from the quantum metric (c.f.~Eq.~(\ref{eq:wf_spread_qgt}) and panel e). (\textbf{d}) Band structure for different values of $V/t$. The orbital character of the X and graphene atoms (A+B) are colored in red and blue, respectively. (\textbf{e}) Quantum metric $\Tr g(\bm{k}) = g_{xx}(\bm{k}) + g_{yy}(\bm{k})$ of the band crossing the Fermi level for $V/t=0.5$ (left) and $V/t=3.0$ (right). Note the different magnitude of scales. The middle panels show corresponding Wannier functions which change the maxima from the X sites ($V/t=0.5$) to the B sites ($V/t=3.0$) by increasing $V$.}
    \label{fig1}
\end{figure}

\prlsection{Hybridization control of flat bands: Quantum geometry and localization}
We start with an elemental model, inspired by early works on functionalized graphene~\cite{Katsnelson2020,Yazyev2007,Boukhvalov2008,Wehling2009,Zhou2009,Yazyev2010,Wehling2010,Hong2012,Nair2012,McCreary2012,Rudenko2013,Mazurenko2016,GonzalezHerrero2016,Li2023} as well as general considerations for flat band systems like Refs.~\cite{Calugaru2021,Regnault2022,Neves2024,Duan2024,Bradlyn2017,Cano2018}. We consider a bipartite lattice with sublattices A and B and one atom residing in each sublattice. Furthermore, sublattice A is assumed to be decorated with impurity atoms called X atoms. All atom species provide an orbital which transforms trivially under local point group operations. Due to the additional impurity, sublattice sites A and B belong to distinct Wyckoff positions of the crystallographic symmetry group. In Fig.~\ref{fig1}a, we illustrate this setup for the example of a decorated two-dimensional honeycomb lattice, i.e., graphene with an impurity X. This geometry belongs to the wallpaper group $p3m1$ (space group 156) with distinct high-symmetry Wyckoff positions 1a, 1b, and 1c on the sublattice sites and hexagon center, respectively, as shown in panel Fig.~\ref{fig1}b.

The single-particle Hamiltonian generically reads $H_0 = \sum_{\bm{k}\sigma}\Psi^\dagger_{\bm{k}\sigma} h(\bm{k}) \Psi_{\bm{k}\sigma}$ with $\Psi_{\bm{k}\sigma} = (c_{{\rm A}\bm{k}\sigma}, c_{{\rm B}\bm{k}\sigma}, c_{ {\rm X}\bm{k}\sigma})^T$ where $c_{m\bm{k}\sigma}$ is the annihilation operator for site $m \in \{\mathrm{A,B,X}\}$, crystal momentum $\bm{k}$, and spin $\sigma$. $h(\bm{k})$ is given by
\begin{equation}
    h(\bm{k})= \begin{bmatrix}
0 & \xi(\bm{k}) & V \\
\xi^*(\bm{k}) & 0 & 0 \\
V^* & 0 & 0 \\
 \end{bmatrix},
 \label{eq:decorated_model}
\end{equation}
where $\xi(\bm{k})$ is the Fourier transform of the hopping $t$ between A and B atoms and $V>0$ the hopping between the X and the A atoms (c.f.~Fig.~\ref{fig1}a). For this Hamiltonian at half-filling, a zero-energy eigenstate $\ket{M,\bm{k}} = C_{\bm{k}}(0,V,-\xi(\bm{k}))^T$ with $C_{\bm{k}}=1/\sqrt{|\xi(\bm{k})|^2+|V|^2}$ exists at any $\bm{k}$, independent of additional system specifics. Only X and B sites contribute to this flat band with the orbital weight being tuned by the hybridization. Furthermore, for any finite $V$ there is a finite gap separating the zero-energy band from the two other bands at $\varepsilon_\pm(\bm{k})=\pm\sqrt{|\xi(\bm{k})|^2+|V|^2}$. This is shown for the decorated honeycomb lattice in Figs.~\ref{fig1}c and \ref{fig1}d where the band structure with orbital/site-weight is shown for different values of $V$.

In the limit of dominating hybridization $V\to\infty$, the flat band's eigenstate becomes $\ket{M,\bm{k}}\to(0,1,0)^T$, i.e., it is entirely localized on the non-decorated B sublattice. This is understandable from a strong bonding-antibonding splitting between $\ket{\mathrm{X}}$ and $\ket{\mathrm{A}}$ states that decouple from the system and push all the low-energy spectral weight into sublattice B. The corresponding Wannier function of $\ket{M,\bm{k}}=(0,1,0)^T$ is centered and sharply peaked on the B site, thus yielding a vanishing Wannier spread.

In the opposite limit $t\gg V\to 0^+$, the impurity atoms decouple and constitute the zero-energy band
$\ket{M,\bm{k}}\to (0,0^+,-\xi(\bm{k})/|\xi(\bm{k})|)^T$ in the entire Brillouin zone except for the nodal point(s) where $\xi(\bm{k})=0$ (c.f.~Fig.~\ref{fig1}d). The flat-band weight is correspondingly located at the impurity atoms. However, since there is no gap closure upon changing the hybridization while keeping it finite ($V>0$), the Wannier center \emph{must not} have moved away from the B site. Conversely, while the Wannier center remains at B, the maxima shift to the three neighboring X sites, spreading over several atoms (c.f.~insets in Fig.~\ref{fig1}e).

The variation in the Wannier spread is not an artifact but required by the quantum geometry of the zero-energy band: The gauge independent quadratic spread of Wannier functions, $\Omega_{\mathrm{I}}$, is determined by the integral of the quantum metric~\cite{Marzari1997,Resta1999,Marzari2012,Peotta2015,Marrazzo2019,Yu2025,SM}\nocite{Provost1980,Mitscherling2025,Profe2024,Resta2011,Yu2025,Marzari1997,Marzari2012,Peotta2015,Blount1962,Katsnelson2020,ryee2023,dzyaloshinskii2003,Wagner2023,Kresse1993,Kresse1996,Kresse1996b,Perdew1996,Bloechl1994,Kresse1999,Grimme2010,Ghosal2025,Mattausch2007,Scheidgen2023_NOMAD,DFT_at_NOMAD,Li2013,Glass2015,wehling2011,schueler2013,kotliar2006,held2007,karolak2010}
\begin{align}
	\Omega_{\mathrm{I}} = \frac{1}{N_{\bm{k}}} \sum_{\bm{k}} g_{xx}(\bm{k}) + g_{yy}(\bm{k}) \leq \langle \bm{r}^2\rangle - \langle \bm{r}\rangle^2
	\label{eq:wf_spread_qgt}
\end{align}
which for single-bands measures the spread of maximally localized Wannier functions~\cite{Marzari2012}. Here, the quantum metric $\Tr g(\bm{k}) = g_{xx}(\bm{k})+g_{yy}(\bm{k})$~\cite{SM}
of the flat band is peaked at the nodes of $\xi(\bm{k})$ for any finite $V>0$. By increasing $V$, it ``leaks out'' by reducing in magnitude and being spread throughout the Brillouin zone, see Fig.~\ref{fig1}e. Accordingly, the minimal spread $\Omega_{\mathrm{I}}$ decreases with increasing $V$ and vanishes for $V\to\infty$, concomitant to the reduction in X-orbital weight in the flat band as shown in Fig.~\ref{fig1}c. Interestingly, $\Omega_{\mathrm{I}}$ is non-monotonic and displays a maximum for small $V$, despite the quantum metric at the nodes increasing monotonically as $V\to0^+$ ($g(\bm{k}=\bm{K})\sim 1/V^2$ for the honeycomb lattice in Fig.~\ref{fig1}~\cite{SM}). 

This structure of localized orbitals hybridizing with a continuum of Dirac states and producing flat bands with a quantum metric peaked at specific spots in the Brillouin zone is not unique to functionalized graphene. It similarly occurs in graphene moiré systems, as highlighted by the topological heavy fermion description of magic-angle twisted bilayer graphene~\cite{Song2022,HerzogArbeitmann2024TopHF}.

\prlsection{Symmetry-distinct site-selective Mott states}
We now investigate how Coulomb interactions conspire with the quantum geometry for the decorated honeycomb lattice. We assume a local Hubbard interaction $U$ acting on all the atomic species:
\begin{align}
 H_\mathrm{int} = U\sum_{m} n_{m\uparrow}  n_{m\downarrow}.
\end{align}	
Here, $\uparrow, \downarrow$ denote the electron spin and $n_{m\uparrow (\downarrow)}$ the number operator. We set the effective local interaction $U=1.6t$ \cite{wehling2011,schueler2013}, which is well below the critical $U$ for the Mott transition of the undecorated graphene \cite{tran2009}. For vanishing hybridization ($V=0$), graphene is a weakly correlated metal and the flat band of purely X character is a Mott insulator due to the lack of kinetic energy. 

The interacting lattice model, $\mathcal{H}=H_0 + H_\mathrm{int}$, is solved within dynamical mean-field theory (DMFT) \cite{DMFT}. The DMFT requires us to solve three independent impurity actions each for A, B, and X sites per self-consistency loop. We solve them using the numerically exact hybridization-expansion continuous-time quantum Monte Carlo method \cite{CTQMC,comctqmc}. 
The self-energy on the Matsubara axis is analytically continued to real frequencies using the maximum entropy method~\cite{Jarrel1996,Bergeron2016}.

\begin{figure}[!tbp]	 
\includegraphics[width=1.0\columnwidth]{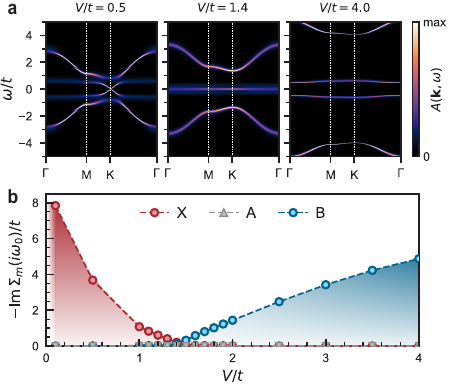}
\caption{Interacting decorated honeycomb model. (\textbf{a}) Momentum-resolved spectral functions from DMFT at $T/t=0.025$ and $U/t=1.6$. From left to right for hybridization $V/t=0.5$, $V/t=1.4$, and $V/t=4$. (\textbf{b}) The imaginary part of the local self-energy at the lowest Matsubara frequency of X (red), A (gray), and B (blue) as function of $V$.}
\label{fig2}
\end{figure}

We present the momentum-resolved spectral functions obtained from DMFT in Fig.~\ref{fig2}a for three representative regimes of hybridization strength. 
In both the small and large $V$ regime [left and right panels], the spectra exhibit clear upper and lower Hubbard bands forming Mott gaps ($\sim U$) with a flat-band metallic phase intervening at intermediate $V$ [middle panel]. Analyzing the underlying self-energy and spin-spin correlation function~\cite{SM} reveals that X and B atoms distinctly drive the aforementioned Mott\-ness and concomitant formation of local spin moments in the small and large $V$ regimes, respectively. Under particle-hole symmetry, this is captured by the imaginary part of the local self-energy $\mathrm{Im}\,\Sigma(i\omega_0)$ at the lowest fermionic Matsubara frequency $\omega_0=\pi k_{\mathrm{B}}T$ ($T$ is temperature), because $\mathrm{Im}\Sigma$ diverges as $\omega_n \to 0$ on the X (B) site for small (large) $V$ as shown in Fig.~\ref{fig2}b. We note that the Dirac cone at $K$ for small $V$ arises from the vanishing X character of the flat band at this point (c.f.~Fig.~\ref{fig1}b), which is otherwise gapped out everywhere else in the Brillouin zone due to the diverging $\mathrm{Im}\Sigma_{\mathrm{X}}(i\omega_n \to 0)$

To understand this behavior, we focus on the extreme limits $V \to 0^+$ and $V \to \infty$, where the spectral weight of the non-interacting flat band predominantly resides on X and B sites, respectively. In these limits, the local self-energy is well approximated by the Hubbard-I form~\cite{hubbard1963} $\Sigma_m (i\omega_n) = U^2/(4i\omega_n)$ for $m=\mathrm{X}$ at $V \to 0^+$ and $m=\mathrm{B}$ at $V \to \infty$, while $\Sigma$ vanishes on the other orbitals. The Hartree term is absorbed into the chemical potential.

Using the analytic form of the Hubbard-I self-energy, the nature of a topological phase transition is revealed by the Luttinger surface---the locus of $\bm{k}$ where the zero-frequency Green's function $G(\bm{k},\omega=0)$ exhibits zero eigenvalues or, equivalently, where the self-energy diverges at $\omega=0$~\cite{dzyaloshinskii2003}. Here, the Luttinger surface consists of a single band of Green's function zeros, which is purely of X character at $V \to 0^+$ and of B character at $V \to \infty$~\cite{SM}. Since X and B occupy distinct Wyckoff positions (c.f.~Fig.~\ref{fig1}b), the site-selective Mott states belong to distinct irreducible representations, i.e., different atomic limits. Hence, the Luttinger surface cannot be deformed continuously by changing $V$ without changing the number of zero bands or breaking symmetries. Consequently, an intermediate metallic phase as in the middle panel of Fig.~\ref{fig2}a has to emerge where the Luttinger surface vanishes. This behavior closely resembles the topological phase transition of noninteracting bands in the Su-Schrieffer-Heeger model~\cite{su1980,asboth2016}. In our case, however, it is the topological phase transition of the ``Mottness'', i.e., the positions of local spin moments, which is associated with the Green's function zeros (not poles) forming the Luttinger surface~\cite{Sakai2009,Gurarie2011,Manmana2012,Volovik2013,Fabrizio2022,Wagner2023,Zhao2023,Blason2023,Setty2024a,Setty2024,Wagner2024,Bollmann2024}. We note that these observations are robust against longer-range hoppings or on-site terms~\cite{SM}.

\prlsection{Proximity coupling of graphene and obstructed Mottness}
An obvious question concerns the realization of the topological transition between two distinct Mott insulating states in material setups. The decorated honeycomb lattice  discussed in this work has been studied extensively in the context of $sp$-electron magnetism in graphene, which is a highly controversial topic in the literature~\cite{Katsnelson2020,Yazyev2007,Zhou2009,Yazyev2010,Hong2012,Nair2012,McCreary2012,GonzalezHerrero2016,Li2023}. Specifically, single-side covalently bonded impurities, like hydrogen or CH$_3$ molecules~\cite{Zhou2009,Boukhvalov2008,Wehling2009,Wehling2010,Rudenko2013,Mazurenko2016}, on graphene correspond to the large $V$ limit of our model. 

A complementary route of functionalization is given by proximity coupling of graphene sheets with suitable two-dimensional systems. Here, we focus on epitaxial graphene proximitized to group-IV adatoms at 1/3 monolayer coverage adsorbed on semiconducting surface. Without graphene, these surface lattice systems show various correlated phases~\cite{Ganz1991,Carpinelli1996,Carpinelli1997,Profeta2005,Profeta2007, Schuwalow2010,Li2011,Hansmann2013,Li2013,Glass2015,Hansmann2016,Jaeger2018,Nakamura2018,Cao2018a,Wu2020a,Ming2023,Ghosal2025}.

\begin{figure}[!tbp]	 
	\includegraphics[width=1.0\columnwidth]{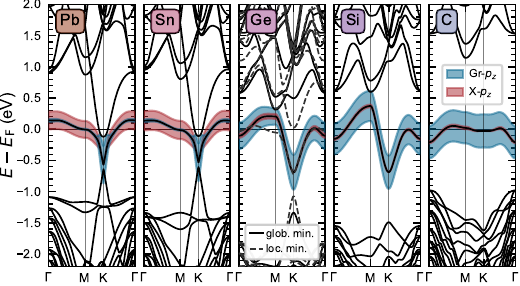}
	\caption{Hybridization control via proximity coupling of epitaxial graphene and group-IV triangular adatom lattices. From left to right, the DFT band structure for different group IV elements X$\in\!\!\lbrace$C,\,Si,\,Ge,\,Sn,\,Pb$\rbrace$ is shown with the weight of the X-$p_z$ and graphene-$p_z$ orbitals as fatband plots. For X\,=\,Ge, two close-in-energy configurations exist (c.f.~Fig.~\ref{fig:DFT_structure} in the End Matter) for which the band structure of the local minimum is additionally drawn with dashed lines and no orbital weight.}
	\label{fig3}
\end{figure}

Motivated by this observation, we investigate the electronic structure of graphene in proximity to different group-IV intercalants (X$\,=\,$C,\,Si,\,Ge,\,Sn,\,Pb) on a SiC(0001) substrate by using density functional theory (DFT)~\cite{SM}. 
The structural details are discussed in the End Matter. The band structure for for each IV element X is drawn in Fig.~\ref{fig3}. They all display a relatively flat band pinned to the Fermi level from the intercalant's $p_z$ orbitals hybridized to the Dirac bands of graphene. Due to the variation in equilibrium distances between graphene and the intercalant atoms, the hybridization strength and orbital character is widely tunable. The site- and orbital-resolved density of states reveals that dominantly X-$p_z$ and B-site graphene-$p_z$ contribute to the formation of the flat band~\cite{SM}.

\begin{figure}[!htbp]	 
	\includegraphics[width=1.0\columnwidth]{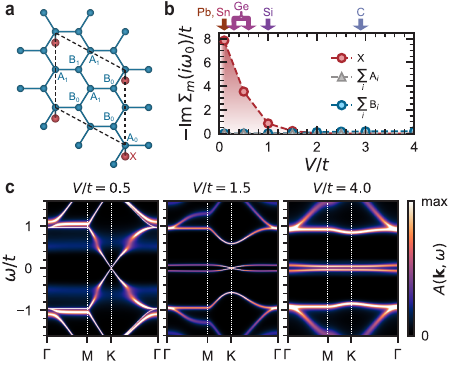}
	\caption{Decorated graphene with smaller impurity density. (\textbf{a}) Unit cell of decorated graphene with 1/8 impurity X coverage. The index of sublattice sites A$_i$, B$_i$ refers to distance from the impurity X. (\textbf{b}) Imaginary part of the local self-energy at the lowest Matsubara frequency of X (red),  cumulative A sites (gray), and cumulative B sites (blue) as function of $V$. Arrows indicate hybridization values estimated for different group IV atoms from fitting the DFT band structures (Fig.~\ref{fig3}). Arrows for X\,=\,Ge at larger and smaller $V$ refer to the local and global energy minimum's configuration, respectively. (\textbf{c}) Momentum-resolved spectral functions from DMFT at $T/t=0.025$ and $U/t=1.6$. From left to right for hybridization $V/t=0.5$, $V/t=1.5$, and $V/t=4$.}
	\label{fig4}
\end{figure}

The band structures obtained from DFT can be rationalized by extending model~(\ref{eq:decorated_model}) to a $2\times2$ unit cell structure with reduced impurity density sketched in Fig.~\ref{fig4}a; model details are given in the End Matter. The resulting spectral functions are shown in Fig.~\ref{fig4}c. The limit of small $V$ ($V/t=0.5$) is very similar to the $1\times 1$ case (c.f.~Fig.~\ref{fig2}a), showing well-formed Hubbard bands originating from the X orbital and a correspondingly large $|\mathrm{Im}\,\Sigma_{\mathrm{X}}(i\omega_0)|\gg t$ (c.f.~Fig.~\ref{fig4}b). Increasing the hybridization quenches Mott-Hubbard correlations at the X site, as indicated by the reduction of $|\mathrm{Im}\,\Sigma_{\mathrm{X}}(i\omega_0)|$ upon increasing $V$ in Fig.~\ref{fig4}b. Likewise, we find a metallic state with a flat band at the Fermi level for intermediate $V/t=1.5$. 

The $V\to\infty$ case is interestingly different in the $2\times 2$ case as compared to the $1\times 1$ case. A pseudogap opens with weak but persisting spectral weight of the flat band at the Fermi level even at $V\gg t$ (Fig.~\ref{fig4}c, right panel), and we do not find a re-entrant divergent component of the self-energy for $V\gg t$ (Fig.~\ref{fig4}b). Instead, the self-energy takes a finite value on several atoms in sublattice B~\cite{SM}.

This qualitative difference originates from the lower density of X sites in the $2\times2$ lattice, as it allows for undecorated A sites and percolating A--B paths (c.f.~Fig.~\ref{fig4}a). While the small-$V$ regime is still dominated by X-site physics, at large $V$ the flat-band spectral weight redistributes over the B sublattice, resulting in a finite quantum metric for all $V>0$ (see End Matter). This geometric obstruction prevents the formation of a (site-selective) Mott state in the strong-hybridization limit. Accordingly, the local self-energy remains finite on the B sites and the system stays a correlated metal with a pseudogap when the flat-band weight predominantly resides on those sites (Fig.~\ref{fig4}c). The hybridization values relevant for the DFT band structures of Fig.~\ref{fig3}, extracted from tight-binding fits~\cite{SM}, span all regimes and are indicated in Fig.~\ref{fig4}b.

The correlated metallic state stabilized at large hybridization due to the finite quantum geometry of the flat band is similar to the vacancy-induced flat band of Ref.~\cite{Marsal2024}. Building on this analogy, disorder or incomplete functionalization may redistribute spectral weight~\cite{Pereira2006,Wehling2007}, but the essential dichotomy between moment formation at weak $V$ and geometrically obstructed metallicity at strong $V$ is expected to persist, as long as there is no prevalent impurity dimerization~\cite{Wehling2007}. Beyond our minimal model, enriching the flat band with multi-orbital degrees of freedom or spin-orbit coupling could qualitatively alter the correlated phases, for instance through Hund's physics or topological Mott states~\cite{WitczakKrempa2014,werner_highspin_2007,demedici_2011,georges_strong_2013,isidori_2019,kugler_2019,ryee_2021}, offering promising directions for future study. A recent theoretical proposal of transition-metal adatom systems with two flat bands and emergent Mott physics~\cite{Menke2024} offers a natural extension of such multi-orbital generalizations via graphene functionalization.

\prlsection{Conclusion}
In this work, we demonstrated a topological phase transition between symmetry-distinct site-selective Mott states and a geometrically stabilized metallic state enabled by hybridization control. We find that in between two states featuring local moments in different Wyckoff positions a correlated metallic state emerges for a most simple model of single-sublattice functionalized graphene. First principles calculations demonstrate that similar physics of a site-selective Mott state transitioning to a geometrically stabilized flat-band metal can be likely realized in intercalated epitaxial graphene platforms. The series of group-IV intercalant atoms realize the entire range from weak ($V\ll t$) to strong hybridization ($V \gg t$).

The models discussed here are indeed very similar to the topological heavy fermion description of magic angle twisted bilayer graphene~\cite{Song2022,HerzogArbeitmann2024TopHF}. In both cases, localized orbitals hybridize with Dirac bands, resulting in a low-energy flat band with substantial quantum geometry. Distinctly, the flat bands in chemically functionalized graphene emerge at chemical energy scales, i.e., scales of covalent bonds and corresponding hoppings $\sim t$. Thus, the correlated flat band physics realized here can persist potentially up to higher temperatures than in twisted bilayer graphene. It appears very promising to explore other possible emergent ordered states, like magnetism and superconductivity, and the corresponding temperature scales in these covalently functionalized graphene systems.

\prlsection{Acknowledgements}
We thank Arka Bandyopadhyay, B.~Andrei Bernevig, Lorenzo Crippa, Stefan Enzner, Andrew J.~Millis, Titus Neupert, and Maia Vergniory for fruitful discussions. This work is funded by the Cluster of Excellence `CUI: Advanced Imaging of Matter' of the Deutsche Forschungsgemeinschaft (DFG) - EXC 2056 - project ID 390715994, by DFG priority program SPP 2244 (WE 5342/5-1, project No.~422707584), the DFG research unit FOR 5242 (WE 5342/7-1, project No.~449119662), the DFG research unit FOR 5249 (``QUAST'', project No.~449872909), and the DFG SFB 1170 Tocotronics (project No.~258499086). J.C.~acknowledges support from the Alfred P.~Sloan Foundation through a Sloan Research Fellowship; from the Flatiron Institute, a division of Simons Foundation; and from the National Science Foundation under Grant No.~DMR-1942447. Calculations were done on the supercomputer Lise at NHR@ZIB as part of the NHR infrastructure under the project hhp00056. The quantum metric was calculated using the divERGe library~\cite{Profe2024}. The DFT data is made available 
at~\cite{DFT_at_NOMAD}.


\bibliography{bib_Mott_top.bib}
{
\onecolumngrid
\section*{End Matter}
\twocolumngrid
}

\begin{figure}[!hb]	 
	\includegraphics[width=1.0\columnwidth]{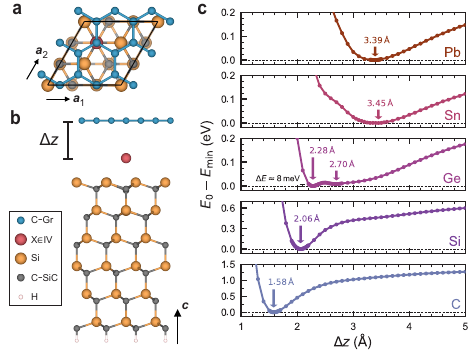}
	\caption{Structure of epitaxial graphene proximitized to group-IV triangular adatom lattices. (\textbf{a}, \textbf{b}) Top and side view of the graphene/X/SiC(0001) heterostructure with distance $\Delta z$ between the graphene layer and the intercalated X atom layer of group IV elements. The X atoms are in a $\sqrt{3}\times\sqrt{3}$ triangular lattice reconstruction on SiC(0001) and graphene is stretched in a $2\times2$ cell. (\textbf{c}) Energy as a function of layer distance $\Delta z$ for different group IV atoms. The minimum energy $E_{\mathrm{min}}$ is taken as zero and the minimum position is marked with an arrow. In case of X\,=\,Ge, two minima close in energy exist.}
	\label{fig:DFT_structure}
\end{figure}

\prlsection{Appendix A: Structure of epitaxial graphene proximitized to group IV adatom monolayer} The general structure of epitaxial graphene in proximity to different
group-IV intercalant monolayers on a
SiC(0001) substrate used in the DFT calculations is shown in Figs.~\ref{fig:DFT_structure}a and \ref{fig:DFT_structure}b. By tuning the intercalant species X$\in\!\!\lbrace$C,\,Si,\,Ge,\,Sn,\,Pb$\rbrace$, the distance $\Delta z$ of the X adatom lattice to the graphene layer is strongly changed in the ground state configuration, as the total energy curves in Fig.~\ref{fig:DFT_structure}c demonstrate. Here, C intercalants are closest to the graphene layer with the equilibrium distance increasing by over 2~\AA~going down the periodic table. Notably, the energy landscape for Ge displays two local minima depending on $\Delta z$. The corresponding band structures of the lowest-in-energy structure for each group IV element X is drawn in Fig.~\ref{fig3}. Technical details of the DFT calculations are given in the Supplemental Material~\cite{SM}.

\begin{figure}[!hb]	 
	\includegraphics[width=1.0\columnwidth]{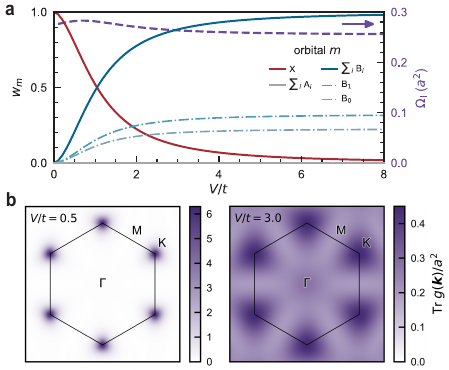}
	\caption{$2\times 2$ graphene supercell with single impurity density. (\textbf{a})  Site-resolved weight $w_m = \sum_{\bm{k}}|w_{\bm{k}m}|^2/N_{\bm{k}}$ ($m\in\lbrace\mathrm{A}_i,\mathrm{B}_i,\mathrm{X}\rbrace$) and minimal Wannier spread $\Omega_{\mathrm{I}}$ of the flat band for the non-interacting $2\time2$ honeycomb supercell model with 1/8 impurity X coverage as sketched in Fig.~\ref{fig4}a. Due to distribution of the flat band weight across B sublattice sites, the Wannier spread $\Omega_{\mathrm{I}}$ does not vanish for $V\to\infty$. (\textbf{b}) Quantum metric in the small (left, $V/t=0.5$) and large (left, $V/t=3.0$) hybridization regime.}
	\label{fig:2x2_model}
\end{figure}

\prlsection{Appendix B: Decorated graphene model with smaller impurity density}
The graphene/X/SiC(0001) heterostructure in Fig.~\ref{fig:DFT_structure}a consists of a $2\times2$ graphene supercell with a single impurity X. This generalization of model~(\ref{eq:decorated_model}) with reduced impurity density of 1/8, illustrated in Fig.~\ref{fig4}a, introduces undecorated A$_1$ sites which connect symmetry-distinct B$_0$ and B$_1$ sites. As a result, the spectral weight of the flat band is distributed across all B sublattice sites as $V$ increases, yielding a robust and finite quantum metric and Wannier spread for all $V>0$, see  Fig.~\ref{fig:2x2_model}a. The momentum dependence of $g(\bm{k})$ also qualitatively changes in the large $V$ limit, whereas the small $V$ regime remains unaffected, since the flat band is dominated by X weight (c.f.~Fig.~\ref{fig:2x2_model}b vs Fig.~\ref{fig1}e).

\input{supple}

\end{document}

%% file: supple.tex
\supplement{Supplemental Material:\\ 
	Real-Space Switching of Local Moments Driven by Quantum Geometry in Correlated Graphene Heterostructures}
\setcounter{secnumdepth}{2}

\makeatletter
\addtocontents{toc}{\string\tocdepth@restore}
\makeatother


\tableofcontents

\section{Quantum metric and bound on Wannier spread}

In this section, we discuss the quantum metric $g_{ij}(\bm{k})$~\cite{Provost1980} in solid-state systems. For this purpose, we concentrate on a periodic Bloch Hamiltonian $h(\bm{k})$ with band energies $\varepsilon_{n\bm{k}}$ and periodic eigenfunctions $\ket{u_{n\bm{k}}}$ of the total Bloch wave function $\ket{\psi_{n\bm{k}}} = \mathrm{e}^{i\bm{k}\hat{\bm{r}}}\ket{u_{n\bm{k}}}$ (with position operator $\hat{\bm{r}}$). The quantum metric can be compactly expressed as
\begin{align}
	g_{ij}(\bm{k}) = \frac{1}{2}\Tr[(\partial_i \hat{P}_{\bm{k}}) (\partial_j \hat{P}_{\bm{k}})]
	\label{eq:quantum_metric}
\end{align}
with $\partial_i\equiv \partial/\partial k_i$ in terms of projector operators
\begin{align}
	\hat{P}_{\bm{k}} = \sum_{n\in\mathcal{B}} \ket{u_{n\bm{k}}}\bra{u_{n\bm{k}}}
	\label{eq:projector_k}
\end{align}
which project to a chosen band manifold $\mathcal{B}$ and which are inherently gauge invariant. We note that working with projectors instead of Bloch wave functions is more advantageous for the analytical and numerical evaluation of geometric invariants like the quantum metric, in particular for numerically implementing the derivative in Eq.~(\ref{eq:quantum_metric})~\cite{Mitscherling2025,Profe2024}.

The quantum metric is the symmetric (real) part of the more general quantum geometric tensor~\cite{Resta2011,Yu2025}
\begin{align}
	Q_{ij} = \Tr[(\partial_i \hat{P})(\mathds{1} - \hat{P})(\partial_j \hat{P})] = \Tr[\hat{P}(\partial_i \hat{P})(\partial_j\hat{P})] = g_{ij} - \frac{i}{2}F_{ij}\,,
\end{align}
with the antisymmetric (imaginary) part being the Berry curvature
\begin{align}
	F_{ij} = i\Tr[\hat{P}(\partial_i\hat{P})(\partial_j\hat{P}) - \hat{P}(\partial_j\hat{P})(\partial_i\hat{P})]\;.
\end{align}
The physical interpretation of the quantum metric $g_{ij}(\bm{k})$ is the measure of distances between two quantum states. For Bloch states, the distance between states at $\bm{k}$ and $\bm{k}+\mathrm{d}\bm{k}$ is given in second order by~\cite{Yu2025}
\begin{align}
	d^2(\bm{k},\bm{k}+\mathrm{d}\bm{k}) = \sum_{ij} g_{ij}(\bm{k})\mathrm{d}k_i\mathrm{d}k_j\;.
\end{align}
While the concept of quantum distances might not be directly tangible~\cite{Provost1980}, recent works have demonstrated that the quantum metric plays a crucial role in various properties -- typically those driven by kinetic effects, such as transport coefficients, spin and superfluid stiffness, and Landau levels -- with its influence being particularly pronounced in flat-band systems~\cite{Yu2025}. A more concrete interpretation for the quantum metric exists in its relationship to the gauge-independent spread of Wannier functions, representing the minimum of the localization functional used to construct maximally localized Wannier functions~\cite{Marzari1997,Marzari2012}. In the following, we demonstrate this connection~\cite{Peotta2015,Yu2025}.

Wannier functions $w_{n\bm{R}}(\bm{r}) = \braket{\bm{r}}{n\bm{R}}$ are the Fourier transform of the Bloch wave functions
\begin{align}
	\ket{n\bm{R}} = \frac{1}{N_{\bm{k}}}\sum_{\bm{k}} \mathrm{e}^{-i\bm{kR}}\ket{\psi_{n\bm{k}}}\quad\mathrm{and}\quad \ket{\psi_{n\bm{k}}} = \sum_{\bm{R}} \mathrm{e}^{i\bm{kR}}\ket{n\bm{R}}.
	\label{eq:Wannier_function_FT}
\end{align}
The localization functional for Wannier functions can be chosen as~\cite{Marzari2012}
\begin{align}
	\Omega = \langle \bm{r}^2\rangle - \langle \bm{r}\rangle^2 = \sum_{n} \left[\bra{n\bm{0}}\hat{\bm{r}}^2\ket{n\bm{0}} - |\bra{n\bm{0}}\hat{\bm{r}}\ket{n\bm{0}}|^2\right] =  \underbrace{\sum_{n}\left[\bra{n\bm{0}}\hat{\bm{r}}^2\ket{n\bm{0}} - \sum_{m\bm{R}}|\bra{n\bm{0}}\hat{\bm{r}}\ket{m\bm{R}}|^2\right]}_{\Omega_{\mathrm{I}}} + \underbrace{\sum_{n,m\bm{R}\neq n\bm{0}}|\bra{n\bm{0}}\hat{\bm{r}}\ket{m\bm{R}}|^2}_{\tilde{\Omega}}
\end{align}
which measures the quadratic spread of Wannier functions around their centers in the unit cell. In the last term, we split the localization functional into terms $\Omega_{\mathrm{I}}$ and $\tilde{\Omega}$ which are both positive definite. The term $\Omega_{\mathrm{I}}$ turns out to be gauge invariant which immediately follows from recasting it in terms of projection operators:
\begin{align}
	\begin{split}
		\Omega_{\mathrm{I}} &= \sum_{n,i} \left[\bra{n\bm{0}}\hat{r}_i^2\ket{n\bm{0}} - \sum_{m\bm{R}}\bra{n\bm{0}}\hat{r}_i\ket{m\bm{R}}\bra{m\bm{R}}\hat{r}_i\ket{n\bm{0}}\right] = \sum_{n,i} \left[\bra{n\bm{0}}\hat{r}_i \left(1 - \sum_{m\bm{R}}\ket{m\bm{R}}\bra{m\bm{R}}\right)\hat{r}_i\ket{n\bm{0}}\right]\\
		&= \sum_{n,i} \left[\bra{n\bm{0}}\hat{r}_i (1 - \hat{P})\hat{r}_i\ket{n\bm{0}}\right]\;,
		\label{eq:Omega_i_wannier_space}
	\end{split}
\end{align}
where we inserted the band-group projector
\begin{align}
	\hat{P} = \sum_{n\bm{R}} \ket{n\bm{R}}\bra{n\bm{R}} \overset{\mathrm{Eq.~(\ref{eq:Wannier_function_FT})}}{=} \sum_{n\bm{k}} \ket{\psi_{n\bm{k}}}\bra{\psi_{n\bm{k}}} = \sum_{n\bm{k}} \ket{u_{n\bm{k}}}\bra{u_{n\bm{k}}} \overset{\mathrm{Eq.~(\ref{eq:projector_k})}}{=} \sum_{\bm{k}} \hat{P}_{\bm{k}}\;.
\end{align}
Hence, minimizing the localization functional $\Omega$ for constructing maximally localized Wannier functions only necessitates the minimization of the noninvariant term $\tilde{\Omega}$. Consequently, $\Omega_{\mathrm{I}}$ constitutes a lower bound to the quadratic spread. We now want to express $\Omega_{\mathrm{I}}$ by Bloch functions. To this end, the matrix elements of $\hat{r}_i$ between Wannier functions become derivatives in momentum space~\cite{Blount1962,Marzari2012}:
\begin{align}
	\bra{n\bm{0}}\hat{r}_i\ket{m\bm{R}} &= \frac{i}{N_{\bm{k}}}\sum_{\bm{k}} \mathrm{e}^{i\bm{kR}} \braket{u_{n\bm{k}}}{\partial_i u_{n\bm{k}}}\;,\\
	\bra{n\bm{0}}\hat{r}^2_i\ket{m\bm{R}} &= -\frac{1}{N_{\bm{k}}}\sum_{\bm{k}} \mathrm{e}^{i\bm{kR}} \braket{u_{n\bm{k}}}{\partial^2_i u_{n\bm{k}}}\;.
\end{align}
Inserting this in Eq.~(\ref{eq:Omega_i_wannier_space}) yields
\begin{align}
	\Omega_{\mathrm{I}} &= \sum_{n,i} \left[\bra{n\bm{0}}\hat{r}^2_i\ket{n\bm{0}} - \sum_{m\bm{R}}\bra{n\bm{0}}\hat{r}_i\ket{m\bm{R}}\bra{m\bm{R}}\hat{r}_i\ket{n\bm{0}}\right] = \frac{1}{N_{\bm{k}}}\sum_{n,i,\bm{k}} \left[-\braket{u_{n\bm{k}}}{\partial^2_i u_{n\bm{k}}} + \sum_{m} \braket{u_{n\bm{k}}}{\partial_i u_{m\bm{k}}}\braket{u_{m\bm{k}}}{\partial_i u_{n\bm{k}}}\right]
	\label{eq:quadratic_spread_calculation_step}
\end{align}
Taking twice the derivative of the orthogonality relation $\braket{u_{n\bm{k}}}{u_{n\bm{k}}} = 1$ gives
\begin{align}
	\braket{\partial_i u_{n\bm{k}}}{\partial_i u_{m\bm{k}}} = -\frac{1}{2}\left(\braket{\partial^2_i u_{n\bm{k}}}{u_{m\bm{k}}} + \braket{u_{n\bm{k}}}{\partial^2_i u_{m\bm{k}}}\right) = -  \braket{u_{n\bm{k}}}{\partial^2_i u_{m\bm{k}}}
\end{align}
since $\braket{u_{n\bm{k}}}{\partial^2_i u_{m\bm{k}}} =  \braket{\partial^2_i  u_{n\bm{k}}}{u_{m\bm{k}}}$ is a real number. Thus, Eq.~(\ref{eq:quadratic_spread_calculation_step}) becomes
\begin{align}
	\Omega_{\mathrm{I}} &= \frac{1}{N_{\bm{k}}}\sum_{n,i,\bm{k}} \left[\braket{\partial_i u_{n\bm{k}}}{\partial_i u_{n\bm{k}}} + \sum_{m} \braket{u_{n\bm{k}}}{\partial_i u_{m\bm{k}}}\braket{u_{m\bm{k}}}{\partial_i u_{n\bm{k}}}\right]\notag\\
	&= \frac{1}{N_{\bm{k}}}\sum_{i,\bm{k}} \frac{1}{2}\Tr[(\partial_i \hat{P}_{\bm{k}})(\partial_i \hat{P}_{\bm{k}})] = \frac{1}{N_{\bm{k}}}\sum_{i,\bm{k}} g_{ii}(\bm{k}) = \frac{1}{N_{\bm{k}}}\sum_{\bm{k}}\Tr g(\bm{k})
\end{align}
showing that $\Omega_{\mathrm{I}}$ and thus the lower bound of the quadratic Wannier spread is determined by the quantum metric.

\subsection{Quantum metric of hybridized flat band model}
The quantum metric of the flat band discussed in the main text is peaked at the nodal points of the dispersion $\xi(\bm{k})$ stemming from the bipartite lattice system. By increasing the hybridization strength $V$ between the impurity X and the sublattice site A, the magnitude of the quantum metric is reduced and becomes distributed over the Brillouin zone. To calculate the quantum metric, we write the projector of the flat band for the model in Eq.~(1) of the main text ($\ket{M,\bm{k}} = (0,V,-\xi(\bm{k}))^T/\sqrt{|\xi(\bm{k})|^2+|V|^2}$):
\begin{align}
	\hat{P}_{\bm{k}} = \ket{M,\bm{k}}\bra{M,\bm{k}} = \frac{1}{|\xi(\bm{k})|^2 + |V|^2}\begin{pmatrix}
		0 & 0 & 0\\
		0 & |V|^2 & -V\xi^*(\bm{k})\\
		0 & -V^*\xi(\bm{k}) & |\xi(\bm{k})|^2
	\end{pmatrix}\;.
\end{align}
The expression of the derivatives $\partial_i \hat{P}_{\bm{k}}$ is lengthy but it becomes significantly simplified at the nodal points. For the case of the honeycomb lattice, the dispersion takes the form $\xi(\bm{q}) = \hbar v[q_x \mp iq_y]$ close to the K and K$^\prime$ points with momentum $\bm{q}=\bm{k}-\bm{K}^{(\prime)}$ and electron velocity $v = 3a|t|/2$ ($a$: lattice constant; $t$: hopping) at the conical points~\cite{Katsnelson2020}. The derivative of the projector at the nodal points is then given by
\begin{align}
	\partial_x \hat{P}_{\bm{k}}\big\vert_{\bm{k}=\bm{K}} = \begin{pmatrix}
		0 & 0 & 0\\
		0 & 0 & -\frac{\hbar v}{V^*}\\
		0 & -\frac{\hbar v}{V} & 0
	\end{pmatrix}\quad,\quad\partial_y \hat{P}_{\bm{k}}\big\vert_{\bm{k}=\bm{K}} = \begin{pmatrix}
		0 & 0 & 0\\
		0 & 0 & i\frac{\hbar v}{V^*}\\
		0 & -i\frac{\hbar v}{V} & 0
	\end{pmatrix}
\end{align}
and we obtain for the quantum metric
\begin{align}
	\Tr g(\bm{K}) = g_{xx}(\bm{K}) + g_{yy}(\bm{K}) = \frac{2\hbar^2v^2}{|V|^2} = \frac{9a^2\hbar^2|t|^2}{2|V|^2}
\end{align}
Thus, the quantum metric at the nodal points diverges as $\propto 1/V^2$ for small hybridization $V>0$ while the Wannier spread $\Omega_{\mathrm{I}}$ remains finite. Note that $g(\bm{k})=0$ for $V=0$ since the lattice model has a different point group symmetry.

\section{Local self-energies and spin-spin correlation functions}
\begin{figure}[!tbp]
	\centering
    \includegraphics[width=0.98\columnwidth]{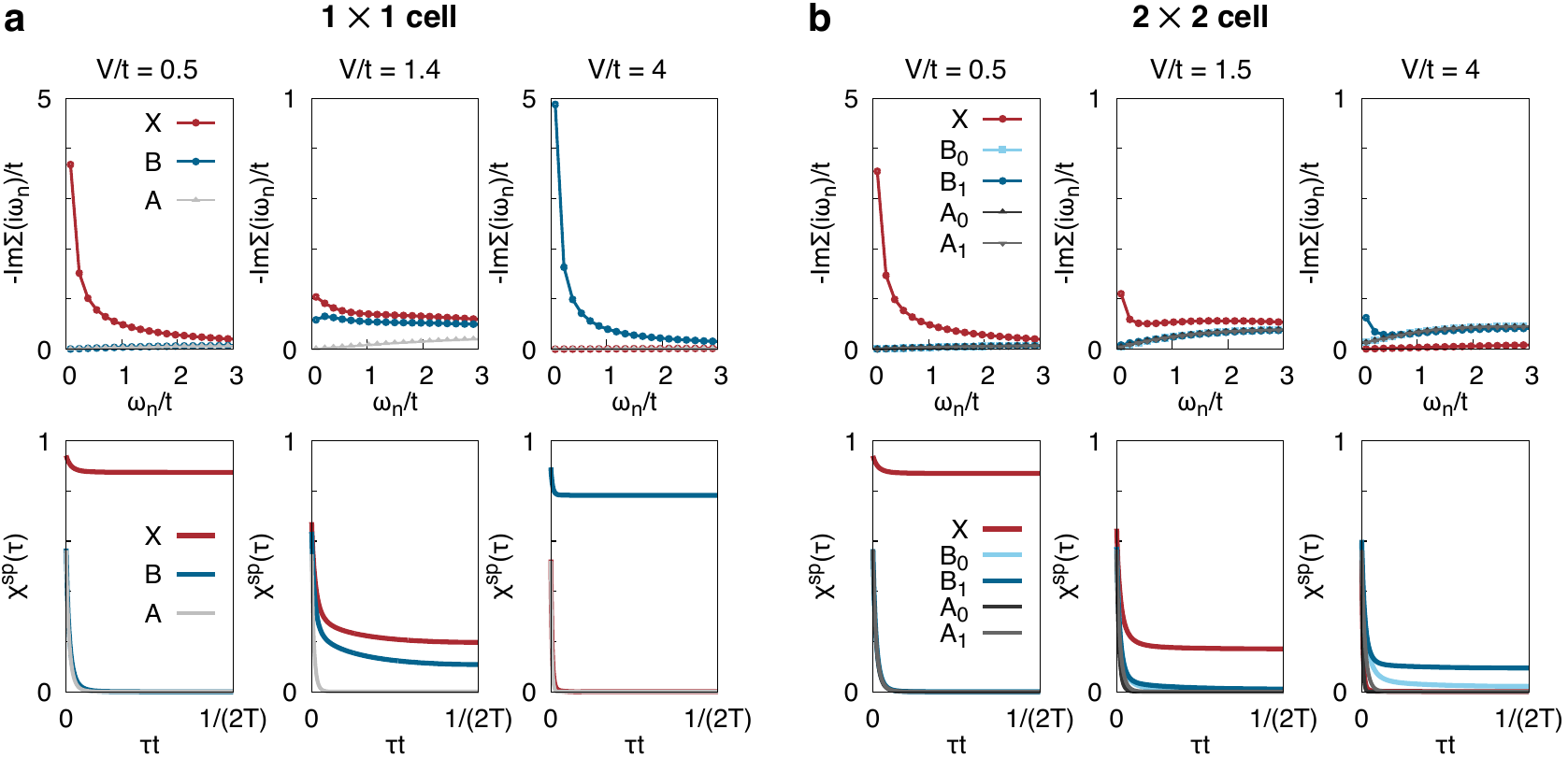}
	\caption{The imaginary part of the local self-energy $\Sigma(i\omega_n)$ on the Matsubara frequency $\omega_n$ axis (top row) and the local spin-spin correlation function $\chi^\mathrm{sp}(\tau)$ on the imaginary time $\tau$ axis (in units of $\mu_{\mathrm{B}}^2$, bottom row) for ({\bf a}) the $1 \times 1$ cell and ({\bf b}) the $2 \times 2$ cell obtained from DMFT at $T/t=0.025$ and $U/t=1.6$.}
	\label{fig_S1_sig}
\end{figure}

\begin{figure}[!htbp]
	\centering
	\includegraphics[width=0.7\columnwidth]{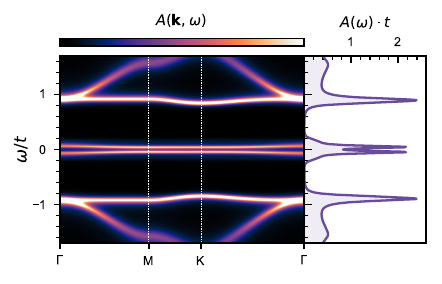}
	\caption{Momentum-resolved $A(\bm{k},\omega)$ and local spectral function $A(\omega) = \sum_{\bm{k}} A(\bm{k},\omega)/N_{\bm{k}}$ for the large hybridization limit ($V/t=4.0$) in the $2 \times 2$ cell model, obtained from DMFT at $T/t=0.025$ and $U/t=1.6$. The left panel is identical to the right panel of Fig.~4c in the main text. In the local spectral function, finite spectral weight exists at $\omega=0$, indicating the occurrence of a pseudogap.}
	\label{fig_S2_locDOS}
\end{figure}

We present in Fig.~\ref{fig_S1_sig} the imaginary part of the local self-energy $\mathrm{Im}\,\Sigma(i\omega_n)$ on the Matsubara frequencies $\omega_n = (2n+1)\pi k_{\mathrm{B}}T$ ($T$: temperature) and the local spin-spin correlation function $\chi^\mathrm{sp}(\tau) = g^2\langle S_z(\tau)S_z \rangle$ (with $S_z(\tau) \equiv (n_\uparrow(\tau) - n_\downarrow(\tau))/2$) on imaginary time $\tau$ obtained from DMFT for three different regimes of $V$, each for the $1\times1$ cell [Fig.~\ref{fig_S1_sig}a] and the $2\times2$ cell [Fig.~\ref{fig_S1_sig}b]. For weak hybridization ($V/t=0.5$), X sites are responsible for the site-selective Mottness and the formation of local spin moments for both the $1 \times 1$-cell and the $2 \times 2$-cell cases. This is highlighted by the divergent self-energies and long-lived unscreened $\chi^\mathrm{sp}(\tau)$, namely $\chi^\mathrm{sp}(1/(2T)) \simeq \chi^\mathrm{sp}(0)$. While the intermediate hybridization regime hosts trivial metallic phases in both cases (middle panels in Figs.~\ref{fig_S1_sig}a and b), the large $V$ regime ($V/t=4$) demonstrates clear disparities between the two models. Only in the $1 \times 1$-cell case, B sites host Mott states (as indicated by the divergent self-energy) and concomitant local spin moments (as indicated by the long-lived unscreened $\chi^\mathrm{sp}(\tau)$). The $2\times2$ cell, on the other hand, does not show (strong) divergent behavior as the weight is distributed over different $B$ sites. Only the B$_1$ site shows a slight upturn of $-\mathrm{Im}\,\Sigma$ at low frequencies, indicating the formation of a pseudogap. This observation is corroborated by the local spectral function $A(\omega) = \sum_{\bm{k}}A(\bm{k},\omega)$ for $V/t=4.0$ shown in Fig.~\ref{fig_S2_locDOS}, which clearly shows a dip but finite spectral weight at the Fermi level. The fate of the pseudogap at lower temperatures remains to be studied.

\begin{figure}[!htbp]
	\centering
	\includegraphics[width=0.7\columnwidth]{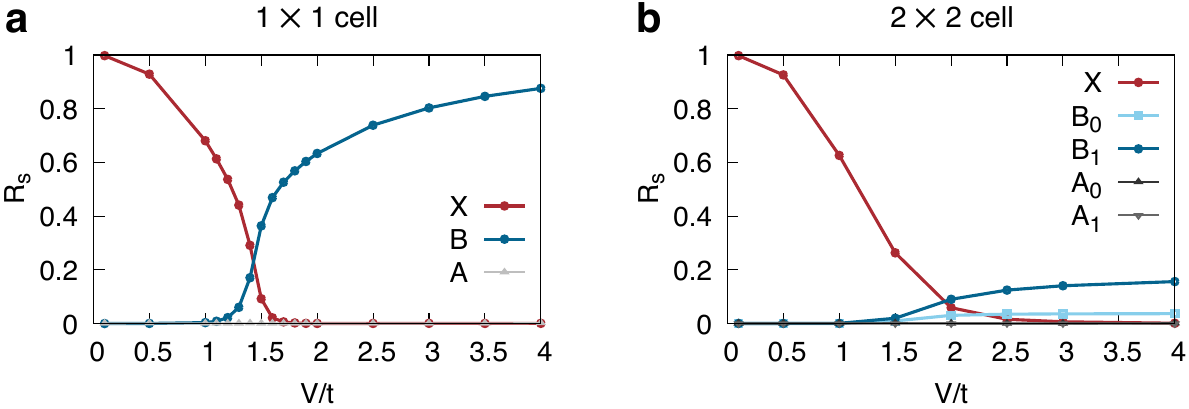}
	\caption{The ``frozen spin ratio'' $R_s$ for ({\bf a}) the $1 \times 1$ cell and ({\bf b}) the $2 \times 2$ cell obtained from DMFT at $T/t=0.025$ and $U/t=1.6$. }
	\label{fig_S3_Rs}
\end{figure}

The local moment formation and their spatial reorganization are also captured by the ``frozen spin ratio'' $R_{\mathrm{s}} \equiv \chi^\mathrm{sp}(1/(2T))/\chi^\mathrm{sp}(0)$ which is a proxy for the degree of spin-Kondo screening \cite{ryee2023}. This quantity is normalized and lies in-between two extreme limits of a fully spin-screened regime ($R_s \to 0$) and the unscreened local moment regime ($R_s \to 1$). Indeed, the hybridization dependence of $R_s$ in Fig.~\ref{fig_S3_Rs} demonstrates the aforementioned dichotomy between the two cells: In the $1 \times 1$ cell, the local spin moment is well-formed in two extreme limits, namely $V/t \to 0^+$ on X and $V/t \to \infty$ on B [Fig.~\ref{fig_S3_Rs}a], whereas it only sets in at $V/t \to 0^+$ on X in the $2 \times 2$ cell [Fig.~\ref{fig_S3_Rs}b].

\section{Green's function zeros and Luttinger surface}
\begin{figure}[!htbp]
	\centering
	\includegraphics[width=1.0\columnwidth]{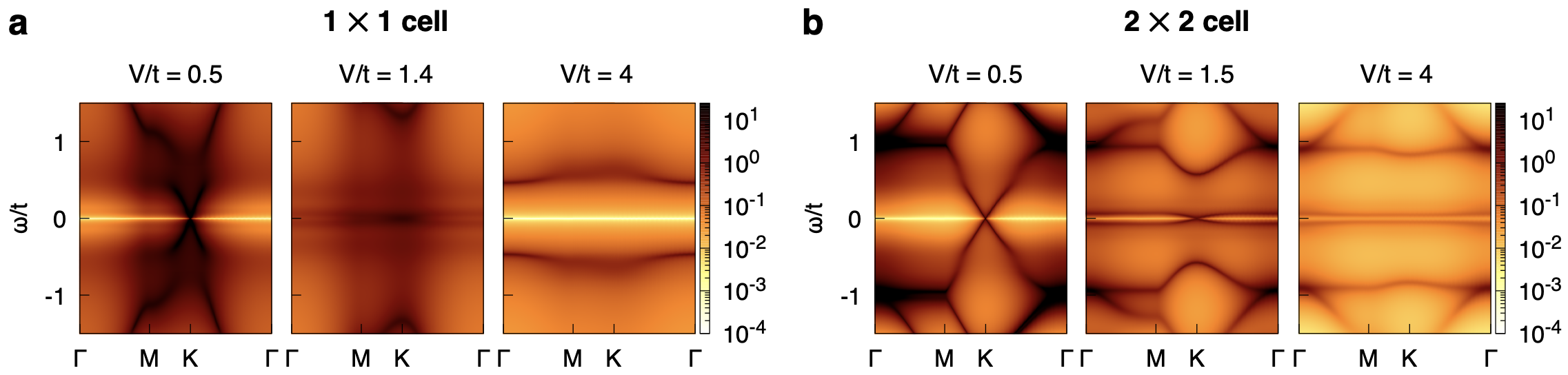}
	\caption{The determinant of the Green's function $|\mathrm{Det}\,G(\bm{k},\omega)|$ highlighting zeros (white) and poles (black). Results are shown for ({\bf a}) the $1\times 1$ and ({\bf b}) the $2\times 2$ cell models obtained from DMFT at $T/t=0.025$ and $U/t=1.6$.}
	\label{fig_S4_detG}
\end{figure}

The Luttinger surface is the locus of $\bm{k}$ where the zero-frequency Green's function $G(\bm{k},\omega=0)$ exhibits zero eigenvalues or, equivalently, where the self-energy diverges at $\omega=0$~\cite{dzyaloshinskii2003}. The analysis of the Green function zeros (GFZs) thereby allows us to characterize the origin of Mottness. To visualize the GFZs, we plot the spectral representation of the determinant of $G(\bm{k},\omega)$ in Fig.~\ref{fig_S4_detG}, which is obtained from analytically continuing the DMFT self-energy to the real axis. In this representation, both zeros (in white) and poles (in black) are highlighted in a combined manner~\cite{Wagner2023}.

Fig.~\ref{fig_S4_detG}a clearly shows a flat band of GFZs (the emergence of the Luttinger surfaces) for both small and large $V$ regimes ($V/t=0.5$ and $V/t=4$) in the $1 \times 1$ cell. The K point is special in this case, as poles and zeros coexist at this point (see discussion of Eqs.~(\ref{eq_GFZ}) and (\ref{eq:GF_zero_eigenvalue_problem}) below). The $2\times2$ cell [Fig.~\ref{fig_S4_detG}b], on the other hand, only displays GFZs in the small $V$ regime ($V/t=0.5$). The absence of GFZs in the large $V$ limit underlines the formation of a pseudogap. This is consistent with our discussion in the main text and the previous section, namely observing a quantum-geometrical obstruction of Mottness for the large $V$ regime in the $2 \times 2$.

We now inspect the nature of the hybridization-tuned topological phase transition between symmetry-distinct site-selective Mott states in the $1 \times 1$-cell case more closely. To this end, we investigate the analytical expression of $\mathrm{Det}\,G(\bm{k},0)$ given by
\begin{align} \label{eq_GFZ}
	\begin{split}
		\mathrm{Det}\,G(\bm{k},0) &= \mathrm{Det} \big[ (\omega\bm{I} - \bm{h}(\bm{k})  - \bm{\Sigma}(\omega))^{-1} \big]\big|_{\omega=0} \\
		&= \frac{1}{ \mathrm{Det}[\omega\bm{I} - \bm{h}(\bm{k})  - \bm{\Sigma}(\omega)]}\big|_{\omega=0}\\
		&= \frac{1}{ |\xi(\bm{k})|^2 \Sigma_\mathrm{X}(0) + |V|^2 \Sigma_\mathrm{B}(0) -\Sigma_\mathrm{A}(0) \Sigma_\mathrm{B}(0)  \Sigma_\mathrm{X}(0)  }\,.
	\end{split}
\end{align}	
Here, $\bm{I}$ is the $3 \times 3$ identity matrix and $\bm{\Sigma}$ the DMFT self-energy matrix, which only has diagonal and momentum-independent components $\Sigma_\mathrm{A}$, $\Sigma_\mathrm{B}$, and $\Sigma_\mathrm{X}$. Using the Hubbard-I approximation ($\Sigma(\omega) = U^2/(4\omega^2)$) as discussed in the main text, it becomes evident from Eq.~(\ref{eq_GFZ}) that $\mathrm{Det}\,G(\bm{k},0)=0$ in the $V/t \to 0^+$ ($V/t \to \infty$) limit due to the diverging $\Sigma_\mathrm{X}(\omega=0)$ ($\Sigma_\mathrm{B}(\omega=0)$). It thus implies that X and B are distinctly responsible for the Mottness in the respective limit.

The real-space switching between symmetry-distinct Mott states is further corroborated by examining the eigenvectors $|\psi(\bm{k}) \rangle_0$ of $G(\bm{k},0)$ for the zero eigenvalues, which satisfy
\begin{align}
	G(\bm{k},0) |\psi(\bm{k}) \rangle_0  = \mathrm{Det}G(\bm{k},0) \cdot \begin{bmatrix}
		\Sigma_\mathrm{B}(0) \Sigma_\mathrm{X}(0) & -\xi(\bm{k})  \Sigma_\mathrm{X}(0) & -V\Sigma_\mathrm{B}(0) \\
		-\xi^*(\bm{k}) \Sigma_\mathrm{X}(0) & \Sigma_\mathrm{A}(0) \Sigma_\mathrm{X}(0) - |V|^2 & \xi^*(\bm{k})V \\
		-V^* \Sigma_\mathrm{B}(0) & \xi(\bm{k}) V^* & \Sigma_\mathrm{A}(0) \Sigma_\mathrm{B}(0) - |\xi(\bm{k})|^2  \\
	\end{bmatrix} |\psi(\bm{k}) \rangle_0 = 0.
	\label{eq:GF_zero_eigenvalue_problem}
\end{align}
Using the Hubbard-I self-energies, we find $|\psi(\bm{k}) \rangle_0 = (0,0,1)^T$ in the $V/t \to 0^+$ limit and $|\psi(\bm{k}) \rangle_0 = (0,1,0)^T$ in the $V/t \to \infty$ limit. Since X and B are on distinct Wyckoff positions 1a and 1b (see Fig.~1b of the main text), the Luttinger surfaces in the two limiting $V/t$ cases belong to two different atomic limits. Hence, the transition between the two site-selective Mott states is accompanied by a topological phase transition of the Luttinger surface.
We note that $|\psi(\tilde{\bm{k}}) \rangle_0$ is ill-defined at the the nodal point (where $\xi(\tilde{\bm{k}})=0$) in the $V/t \to 0^+$ limit. It is in fact a natural consequence of the case where both poles (due to $\xi(\tilde{\bm{k}})=0$) and zeros (due to the divergent $\Sigma_\mathrm{X}(0)$) of $G(\tilde{\bm{k}},0)$ coincide at the same $\tilde{\bm{k}}$ point(s), c.f.~Eq.~(\ref{eq_GFZ}). For this reason, the flat band of zeros is intersected by the poles of Dirac cones in Fig.~\ref{fig_S4_detG} for $V/t=0.5$.

\section{Effects of longer-range hopping and different onsite energies}
\begin{figure}[h]
	\floatbox[{\capbeside\thisfloatsetup{capbesideposition={right,center},capbesidewidth=7cm}}]{figure}[\FBwidth]
	{\caption{Lattice structure of decorated graphene as in Fig.~1 of the main text with additional longer-range hoppings $V'$ (between X and B) and $t'$ (between the same graphene sublattices) and finite onsite energy $\Delta_\mathrm{CF}$ of X.}\label{fig_longrange_model}}
	{\includegraphics[width=5cm]{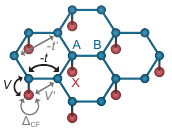}}
\end{figure}

\begin{figure}[!htbp]
	\centering
	\includegraphics[width=0.98\columnwidth]{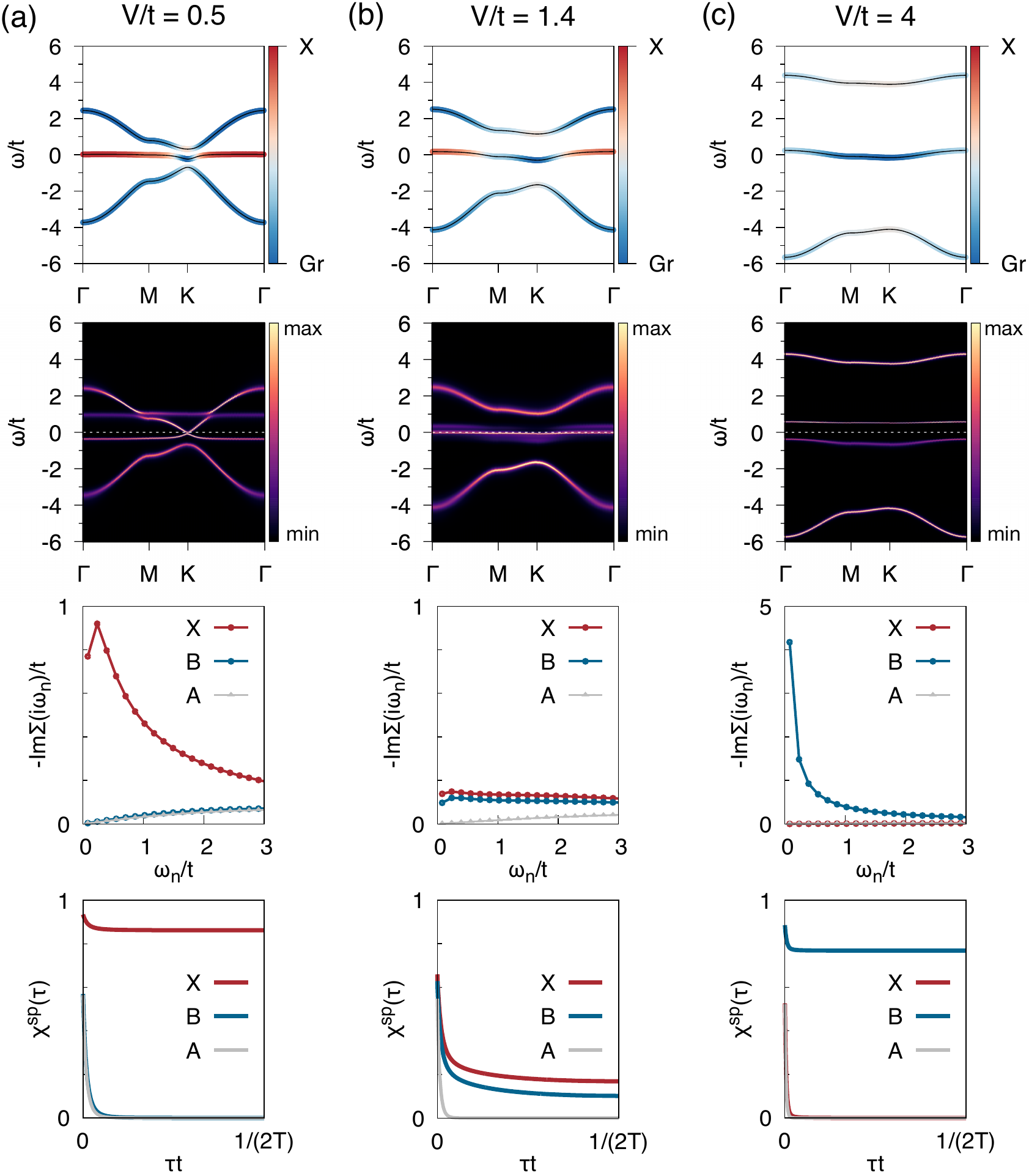}
	\caption{Band structures and DMFT results for the model described in Fig.~\ref{fig_longrange_model}. First row: noninteracting band structures. Second row: momentum-resolved spectral functions obtained from DMFT. White horizontal dotted lines indicate the Fermi level. Third row: imaginary part of the local DMFT self-energy, $\mathrm{Im}\Sigma(i\omega_n)$, as a function of Matsubara frequency $\omega_n$. Fourth row: local spin–spin correlation function, $\chi^\mathrm{sp}(\tau)$, as a function of imaginary time $\tau$.  All results are for $T=0.025t$ and $U/t=1.6$.}
	\label{fig_longrange_DMFT}
\end{figure}

In the main text, we considered the simplest model with only nearest-neighbor hopping amplitudes $t$ and $V$ and identical on-site energies for all atom species. Here, we examine the effects of longer-range hoppings $t'$ and $V'$ and additional on-site energy term $\Delta_\mathrm{CF}$ (crystal field splitting) for X as shown in Fig.~\ref{fig_longrange_model}. The single-particle Hamiltonian now reads (neglecting spin $\sigma$)
\begin{align}
		H_0 = -t \sum_{\langle ij\rangle} c^{\dagger}_{\mathrm{A}i}c^{}_{\mathrm{B}j} -  t^{\prime}\sum_{\langle\!\langle ij\rangle\!\rangle} (c^{\dagger}_{\mathrm{A}i}c^{}_{\mathrm{A}j} + c^{\dagger}_{\mathrm{B}i}c^{}_{\mathrm{B}j} )
		+ V\sum_{i} c^{\dagger}_{\mathrm{A}i}c^{}_{\mathrm{X}i} + V^{\prime}\sum_{i} c^{\dagger}_{\mathrm{B}i}c^{}_{\mathrm{X}i} + \Delta_{\mathrm{CF}}\sum_{i}n_{\mathrm{X}i} + h.c. ,
		\label{eq:longrange}
\end{align}
where $\langle ij\rangle$ and $\langle\!\langle ij\rangle\!\rangle$ denote hopping between nearest- and next-nearest-neighboring sites, respectively. We set $t'=0.036t$ by taking the actual $t'/t$ ratio in graphene~\cite{Katsnelson2020} and $V'=\Delta_\mathrm{CF}=0.2t$. 
The corresponding noninteracting band structures are shown in Fig.~\ref{fig_longrange_DMFT} for three representative values of $V$: $V/t=0.5$, 1.4, and 4. As in the original $t$--$V$ model, spectral weight of the ``flat" band crossing the Fermi level is located at the X site for small $V/t$, whereas at the B site for large $V/t$. The A site has vanishingly small weight in the flat band for the entire $V$ range. Note that due to the additional one-body terms ($t', V', \Delta_\mathrm{CF}$), particle-hole symmetry no longer holds in this case.

Most importantly, we find that the same physics discussed for the simplest model also emerges when longer-range hoppings and finite on-site energies are included (provided they are not too large to substantially modify the noninteracting band structure). As shown in Fig.~\ref{fig_longrange_DMFT}, the DMFT spectral functions (second row) exhibit clear Mottness for both small and large $V/t$, separated by an intermediate metallic state emerging in between. The Mott state in each regime is accompanied by the formation of a local spin moment (fourth row) on the X site for $V/t=0.5$ and on the B site for $V/t=4$, demonstrating a real-space switching of the local spin moment upon tuning $V$. It is worth noting that, unlike in the simplest model discussed earlier, the absence of particle-hole symmetry implies that the imaginary part of the local DMFT self-energy, $\mathrm{Im}\Sigma(i\omega_n)$, does not necessarily diverge as $\omega_n \rightarrow 0$ in the Mott state (cf. third row in Fig.~\ref{fig_longrange_DMFT}).

We finally comment on the effects of nonlocal density-density interactions. In a simplified mean-field (Hartree-Fock) scheme, nonlocal density-density interactions will generate two effects: (i) additional onsite energies due to the Hartree self-energy and (ii) additional hopping amplitudes arising from the nonlocal Fock self-energy. Since these two effects are at the one-body level, they can be effectively incorporated into a ``new'' single-particle Hamiltonian $H_0$ that includes additional long-range hoppings and onsite energies stemming from the mean-field self-energy. As we have demonstrated above, the site-selective Mott physics is robust against longer-range hoppings and additional onsite energies. Thus a mean-field treatment of nonlocal density-density interactions (provided they are not too large) does not alter the site-selective Mott physics. If nonlocal density-density interactions are large enough to effectively screen out the local interaction $U$, the system may undergo a phase transition to a charge-ordered state.

\section{DFT calculations and extraction of hybridization}
\subsection{Numerical details of the DFT calculations}
The DFT calculations were carried out using the Vienna Ab initio Simulation Package (VASP)~\cite{Kresse1993, Kresse1996, Kresse1996b} employing the PBE exchange-correlation functional~\cite{Perdew1996} and the projector-augmented wave (PAW) formalism~\cite{Bloechl1994,Kresse1999}. A plane-wave cutoff of 400~eV, a $\bm{k}$-mesh of $12\times12\times1$, and DFT-D3 van der Waals corrections were used~\cite{Grimme2010}. Structure relaxations were done until forces were smaller than $5$~meV/\AA. To calculate the energy dependence on the distance $\Delta z$ between X atom and graphene layer (c.f.~Fig.~3c in the main text), all atoms were allowed to move except for the $z$ positions of the X atom and the graphene C atom located directly above. The unit cell, consisting of X adatoms on a $\sqrt{3}\times\sqrt{3}$ cell of the SiC(0001) substrate and a (stretched) $2\times2$ graphene (c.f.~Fig.~3a and b of the main text), amounts to a local approximation of the larger $6\sqrt{3}\times6\sqrt{3}$ cell that is realized in experiments~\cite{Ghosal2025,Mattausch2007}. The DFT data for the energetic global minima -- and in the case of X$\,=\,$Ge also the additional local minimum -- are deposited on the NOMAD repository~\cite{Scheidgen2023_NOMAD} at \cite{DFT_at_NOMAD}.

\subsection{Site- and orbital-resolved density of states}
\begin{figure}[!htbp]
	\centering
	\includegraphics[width=1\columnwidth]{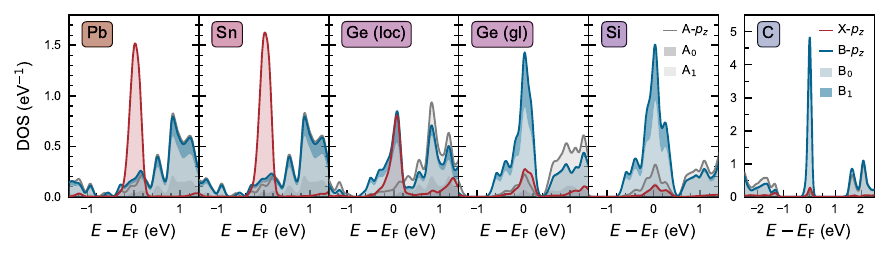}
	\caption{Site- and orbital-resolved density of states (DOS) for the graphene/X/SiC heterostructures for different group IV atoms $\mathrm{X}\in\lbrace$Pb,\,Sn,\,Ge,\,Si,\,C$\rbrace$. For X\,=\,Ge, both local (loc) and global (gl) minimum configuration (c.f.~Fig.~3c in main text) are shown. The DOS of X (red), B (blue), and A (grey) site $p_z$-orbitals are shown with solid lines. The shaded area below the curves indicate contributions from symmetry-distinct A$_0$ (dark grey), A$_1$ (light grey), B$_0$ (light blue), and B$_1$ (dark blue) sites; see Fig.~4a of the main text for the structure.}
	\label{fig_S7_DFT_LDOS}
\end{figure}

The flat band emerging in the graphene/X/SiC(0001) heterostructures at zero energy is formed from the $p_z$ orbitals of either the X or graphene atoms as shown in the DFT band structure in Fig.~3d of the main text. To further elaborate on the dominant contribution to the flat band, we inspect the site-resolved density of states (DOS) in Fig.~\ref{fig_S7_DFT_LDOS} which uses the same site labels as Fig.~4a of the main text. Depending on the group IV atom X, the dominant contribution to the zero-energy flat band either stems from the X-$p_z$ or B-site-$p_z$ orbitals while the A-site contribution remains negligibly small for all structures. Thus, only the sublattice B sites of graphene contribute to the flat band in accordance with our effective modeling.

\subsection{Extraction of hybridization values for the $2\times2$ cell}
The electronic structure of the graphene/X/SiC(0001) heterostructure can be qualitatively captured by a $2\times2$ cell generalization of the decorated honeycomb (Eq.~(1) in the main text), featuring one X atom for every eight graphene atoms. To obtain a \textit{qualitative} assessment of the hybridization $V$ for each group IV element, we fitted the $2\times2$ cell model to the DFT band structure. This fitting utilized an extended model with additional parameters in order to account for the complexities of the real material's structure. The extended model's Hamiltonian reads (neglecting spin $\sigma$)
\begin{align}
		H_0 = t_{\mathrm{Gr}} \sum_{\langle ij\rangle} c^{\dagger}_{\mathrm{A}i}c^{}_{\mathrm{B}j} +  t^{\prime\prime}_{\mathrm{Gr}}\sum_{\langle\!\langle ij\rangle\!\rangle} c^{\dagger}_{\mathrm{A}i}c^{}_{\mathrm{B}j} + t_{\mathrm{X}} \sum_{\langle ij\rangle} c^{\dagger}_{\mathrm{X}i}c^{}_{\mathrm{X}j} +  t^{\prime}_{\mathrm{X}}\sum_{\langle\!\langle ij\rangle\!\rangle} c^{\dagger}_{\mathrm{X}i}c^{}_{\mathrm{X}j}
		+ V\sum_{i} c^{\dagger}_{\mathrm{A}i}c^{}_{\mathrm{X}i} + \Delta_{\mathrm{CF}}\sum_{i}n_{\mathrm{X}i} + \delta \sum_{i\alpha} n_{\alpha i} + h.c.
		\label{eq:extended_model}
\end{align}
where $\langle ij\rangle$ and $\langle\!\langle ij\rangle\!\rangle$ denote hopping between nearest- and next-nearest-neighboring sites, respectively, for the corresponding sublattice sites. Besides the nearest-neighbor hopping $t_{\mathrm{Gr}}\equiv -t$ in graphene and hybridization $V$, the parameters considered include: nearest- and next-nearest-neighbor hopping $t_{\mathrm{X}}$ and $t_{\mathrm{X}}^\prime$ between X atoms; crystal field splitting $\Delta_{\mathrm{CF}}$ between the on-site energy of the X-$p_z$ and graphene-$p_z$ orbitals; global shift $\delta$ of the band structure; and next-nearest-neighbor hopping $t_{\mathrm{Gr}}^{\prime\prime}$ between graphene sublattices A and B (which is an effective third-order hopping in graphene). The next-nearest-neighbor hopping ($t_{\mathrm{Gr}}^\prime$) within the same sublattice was neglected due to its minimal contribution in freestanding graphene~\cite{Katsnelson2020}. 

\begin{figure}[t]
	\centering
	\includegraphics[width=1\columnwidth]{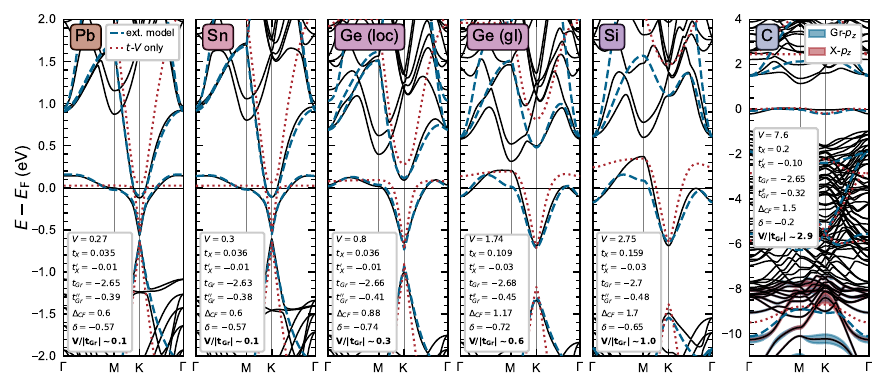}
	\caption{Extended decorated graphene model (Eq.~(\ref{eq:extended_model}), blue dashed lines) fitted to the DFT band structure (black solid lines) of graphene/X/SiC for different group IV atoms $\mathrm{X}\in\lbrace$P,\,Sn,\,Ge,\,Si,\,C$\rbrace$. Additionally, the decorated graphene model with only nearest-neighbor graphene hopping $t\equiv t_{\mathrm{Gr}}$, hybridization $V$, and crystal field splitting $\Delta_{\mathrm{CF}}$ is drawn with red dotted lines. For X\,=\,Ge, both local (loc) and global (gl) minimum configuration (c.f.~Fig.~3c in main text) are shown. For the case of X\,=\,C, fatbands of the Gr-$p_z$ and X-$p_z$ orbital weights are drawn for the high-energy bands around $-10$ to $-8$\,eV, as these arise from large hybridization $V$ and need to be captured in the fitting. Model parameters indicated in the panel insets are given in eV. Note that $\delta$ describes an overall shift of the tight-binding models with respect to the DFT band structure.}
	\label{fig_S8_Vfitting}
\end{figure}

The fitted band structure is shown in Fig.~\ref{fig_S8_Vfitting}. In addition, we show the band structure of a minimal model (red dotted lines) which only contains nearest-neighbor graphene hopping $t_{\mathrm{Gr}}\equiv -t$, hybridization $V$, and crystal field splitting $\Delta_{\mathrm{CF}}$. Overall, the increase of hybridization strength is well captured in both the extended and minimal model  by going from heaviest (Pb) to lightest (C) group IV element. Note that X\,=\,C lies in the large hybridization limit and the value of $V$ was chosen to fit the order of magnitude of the high energy bands with Gr-$p_z$ orbital weight at around $-10$ to $-8$\,eV (see last panel of Fig.~\ref{fig_S8_Vfitting}).

\section{DFT+DMFT of graphene/X/SiC(0001)}
Using the single-particle Hamiltonian constructed for the graphene/X/SiC(0001) heterostructures [Eq.~(\ref{eq:extended_model})], we perform DFT+DMFT calculations to demonstrate that the physics discussed in the $2 \times 2$ cell model is indeed realized in the graphene/X/SiC(0001) systems. To this end, we consider two extreme limits of the series, namely $\mathrm{X}=\mathrm{Pb}$ and $\mathrm{X}=\mathrm{C}$, which correspond to the smallest ($V/t \simeq 0.1$) and largest ($V/t \simeq 2.9$) hybridization strengths, respectively. We adopt $U = 1.4$~eV for the X sites considering the values ranging from $1.2$ to $2$~eV used in previous studies on similar systems \cite{Li2013,Glass2015,Ghosal2025}, and an {\it ab initio} estimate of $U = 4.24$~eV for the graphene sites \cite{wehling2011,schueler2013}. However, our conclusions are robust with respect to reasonable variations in the $U$ values.

An important issue within the DFT+DMFT framework is the choice of the double-counting (DC) self-energy \cite{kotliar2006,held2007,karolak2010}. This term arises because DFT already accounts for part of the many-body interactions through the Hartree and exchange-correlation terms. Consequently, the portion already included in DFT must be subtracted from the local self-energy obtained within DMFT. 
How to specify the DC term remains an important open problem, and no general consensus has been reached. We therefore examine two different DC self-energies:
\begin{align}
    \Sigma_{m\sigma}^\mathrm{DC-1} &= U_{m}n^0_{m\bar{\sigma}} \\
    \Sigma_{m\sigma}^\mathrm{DC-2} &= U_{m}\big( n^0_{m\uparrow} + n^0_{m\downarrow} - \frac{1}{2} \big),
\end{align}
where $n^0_{m\sigma}$ denotes the electron occupation of site $m$ with spin $\sigma$ obtained from the single-particle Hamiltonian Eq.~(\ref{eq:extended_model}). The first prescription (DC-1) assumes that the DFT electronic structure already incorporates the Hartree self-energy of a given interaction Hamiltonian, whereas the second (DC-2) follows the formula suggested in Ref.~\cite{held2007}. As will be shown below, the choice of DC self-energy does not affect our results.

\begin{figure}[!htbp]
	\centering
	\includegraphics[width=1.0\columnwidth]{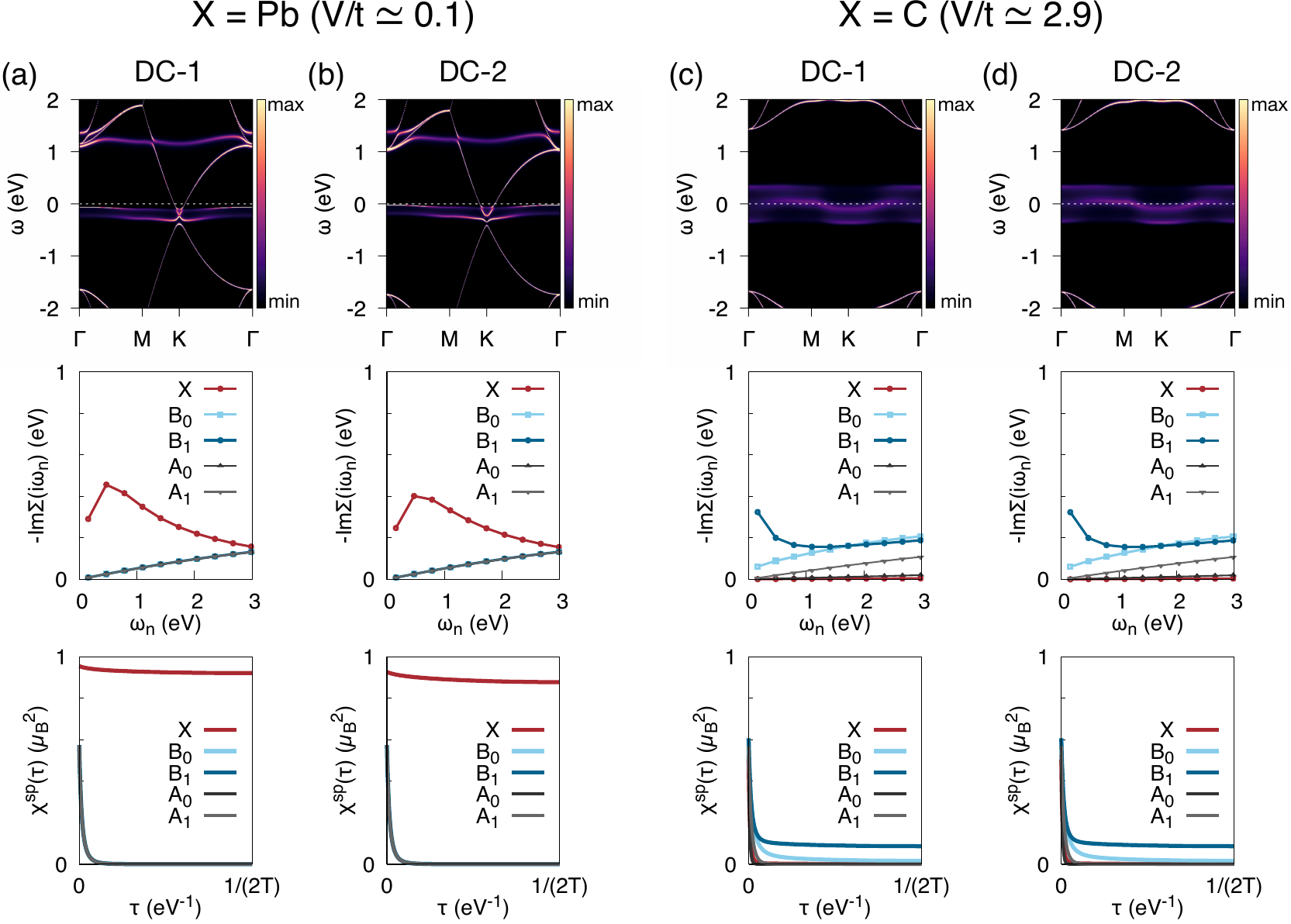}
	\caption{Top row: momentum-resolved spectral functions obtained from DFT+DMFT calculations. Middle row: imaginary part of the local DMFT self-energy, $\mathrm{Im}\Sigma(i\omega_n)$, as a function of Matsubara frequency $\omega_n$. Bottom row: local spin–spin correlation function, $\chi^\mathrm{sp}(\tau)$, as a function of imaginary time $\tau$. All results are for $T=0.05$~eV.}
	\label{fig_DFT+DMFT}
\end{figure}

Our calculation results are shown in Fig.~\ref{fig_DFT+DMFT}. 
The physics discussed for the $2 \times 2$ cell model is indeed captured by the graphene/X/SiC(0001) heterostructures. Specifically, a well-formed local moment appears on the X site ($\mathrm{X}=\mathrm{Pb}$) [Figs.~\ref{fig_DFT+DMFT}a and b] for weak hybridization, whereas in the large-hybridization case ($\mathrm{X}=\mathrm{C}$) [Figs.~\ref{fig_DFT+DMFT}c and d] it is suppressed due to substantial quantum-geometric effects, which hinder Mottness and thus the formation of a strong local spin moment.
As in the case of Fig.~\ref{fig_longrange_DMFT}, $\mathrm{Im}\Sigma(i\omega_n)$ of X for the $\mathrm{X}=\mathrm{Pb}$ does not diverge as $\omega_n \to 0$ since particle-hole symmetry is broken.  
Note also that the large-hybridization case ($\mathrm{X}=\mathrm{C}$) does not exhibit a pseudogap, unlike in Fig.~\ref{fig_S2_locDOS} for the $2 \times 2$ cell model. Understanding whether this is due to the absence of particle-hole symmetry, and how it evolves upon lowering the temperature, requires further investigation.

\vfill

\makeatletter
\addtocontents{toc}{\string\tocdepth@munge}
\makeatother
\makesuppbib
\supplementEnd